\documentclass[aps,prl,twocolumn]{revtex4}
\usepackage{graphicx}
\usepackage{amssymb}
\usepackage[dvips]{color}
\begin{document}
\title{Incoherent matter-wave solitons}
\author{H. Buljan,$^{1,2}$ M. Segev,$^{1}$ and A. Vardi$^{3}$}
\affiliation{ $^{1}$Physics Department, Technion -
Israel Institute of Technology, Haifa 32000, Israel}
\affiliation{
$^2$Department of Physics, University of Zagreb, PP 332, Zagreb, Croatia}
\affiliation{
$^3$Department of Chemistry, Ben Gurion University of Negev, Beer Sheva, Israel}

\date{\today}


\begin{abstract}
The dynamics of matter-wave solitons in Bose-Einstein condensates (BEC) is 
considerably affected by the presence of a surrounding thermal cloud and by 
condensate depletion during its evolution. We analyze these aspects of BEC 
soliton dynamics, using time-dependent Hartree-Fock-Bogoliubov (TDHFB) theory. 
The condensate is initially prepared within a harmonic trap at finite 
temperature, and solitonic behavior is studied by subsequently propagating the 
TDHFB equations without confinement. Numerical results demonstrate the collapse
of the BEC via collisional emission of atom pairs into the thermal cloud, resulting
in splitting of the initial density into two solitonic structures with opposite 
momentum. Each one of these solitary matter waves is a mixture of condensed and 
noncondensed particles, constituting an analog of optical random-phase
solitons.  
\end{abstract}

\pacs{03.75.Lm, 03.75.Be}

\maketitle


The physics of quantum-degenerate, interacting bose gases closely resembles
the behavior of  light in nonlinear media. The dynamics of Bose-Einstein
condensate (BEC) at zero-temperature is well-described by the Gross-Pitaevskii 
(GP) mean-field theory, given by the nonlinear Schr\"odinger equation (NLSE)
for the condensate order parameter. The same equation describes the evolution
of coherent light in nonlinear Kerr media. This analogy has opened the way 
for the field of nonlinear atom optics \cite{Lens,Rolston} with striking 
demonstrations of familiar nonlinear optics phenomena such as four wave mixing 
\cite{Deng}, superradiant Rayleigh scattering \cite{Inouye1}, and matter-wave 
amplification \cite{Kozuma,Inouye2}, carried out with matter-waves.
One such phenomenon is the formation of matter-wave solitons 
\cite{Ruprecht,PerezG,Burger,Denschlag,Busch,Smerzi,Khajkovic,Strecker,Salasnich,
Eiermann}. Experimentally, dark solitons \cite{Burger,Denschlag} and bright 
gap solitons \cite{Eiermann} were observed in BECs with repulsive interactions,
whereas bright solitons  \cite{Khajkovic,Strecker} were demonstrated in systems 
with attractive interactions. These experimental results are augmented by extensive 
theoretical work including predictions on bright \cite{Ruprecht,PerezG} and
dark \cite{Busch} matter-wave solitons, lattice solitons \cite{Smerzi},
and soliton trains \cite{Salasnich}. The vast majority of previous theoretical 
efforts on matter-wave solitons have utilized the zero-temperature GP
mean-field theory. However, in a realistic system, elementary excitations
arising from thermal and/or quantum fluctuations are always present \cite{Griffin}, 
and the BEC dynamics may be considerably affected by the motion of excited atoms 
around it (thermal cloud), and by its dynamical depletion \cite{Castin}, giving rise
to new nonlinear matter-wave phenomena.

Here we analyze these aspects of BEC soliton dynamics by using the time-dependent 
Hartree-Fock-Bogoliubov (TDHFB) theory \cite{Proukakis1,Holland,Rey}. 
Soliton dynamics is analyzed by first calculating the finite-temperature ground state 
of the attractively interacting gas within an harmonic trap \cite{Griffin}. 
The harmonic confinement is then suddenly turned off, and the partially condensed 
Bose gas starts to dynamically evolve. Within the TDHFB model, we find a 
characteristic pattern of evolution of the system whereby pairs of atoms are 
collisionally excited from the BEC into the thermal cloud causing the initial 
density to eventually split into two solitonic structures with opposite momentum.  
Both solitons constitute a mixture of condensed and noncondensed particles.  
We emphasize that the observed composite waves are a truly novel type of 
matter-wave solitons, where localization is attained not only in spatial density, 
but also in spatial correlations. This type of solitons are reminiscent of 
composite incoherent optical solitons  \cite{Mitchell,Equ,Mitchell1}, thus 
highlighting an analogy between incoherent light behavior in nonlinear media 
and BECs at finite-temperatures.

Starting with a near-unity condensate fraction at very low temperatures, 
the GP dynamics reproduces, under proper conditions, the well-known 
zero-temperature BEC solitons, thus demonstrating the condensate's mechanical
stability. However, the evolution of the same initial nearly pure BEC with 
TDHFB clearly illustrates BEC depletion through pairing, causing these 
coherent solitons to disintegrate in a characteristic fashion into incoherent 
solitary matter waves. In all cases, when both the trap and the interparticle 
interactions are turned off simultaneously, we observe fast matter-wave
dispersion.

We consider a system of $N$ interacting bosons placed in a quasi
one-dimensional (Q1D) cigar-shaped harmonic potential
$V_{ext}(x,y,z)=(\omega_x x^2+\omega_{\perp} y^2+\omega_{\perp} z^2)/2$,
where $\omega_{\perp}\gg\omega_x$ denote the transverse and the
longitudinal frequencies of the trap, respectively.
The interparticle interaction is approximated by the Q1D contact potential
$V(x_1-x_2)=g_{1D}\delta(x_1-x_2)$, where $g_{1D}=-2\hbar^2/ma_{1D}$,
$a_{1D}\approx -a_{\perp}^2/a_{3D}$ is the effective 1D scattering 
length \cite{Dunjko,Kheruntsyan1D,Moritz}, $m$ is the particle mass, 
$a_{\perp}=\sqrt{\hbar/m\omega_{\perp}}$ is the size
of the lowest transverse mode, while $a_{3D}$ is the 3D scattering length.
At finite temperatures, the equilibrium state of the system can be described 
by the HFB theory \cite{Griffin}:

\begin{equation}
H_{sp} \Phi^{s}+
g_{1D}[n_c^{s}(x)+2\tilde n^{s}(x)]\Phi^{s}
+g_{1D} \tilde m^{s}(x)\Phi^{s*}
=\mu \Phi^{s}(x),
\label{statcon}
\end{equation}
\begin{equation}
\left [
\begin{array}{c c}
\mathcal{L}^{s}(x) &
\mathcal{M}^{s}(x) \\
-\mathcal{M}^{s^*}(x) & 
-\mathcal{L}^{s}(x)
\end{array}
\right ]
\left [
\begin{array}{c}
u_j^{s}(x) \\
v_j^{s}(x) 
\end{array}
\right ]
=E_j
\left [
\begin{array}{c}
u_j^{s}(x) \\
v_j^{s}(x) 
\end{array}
\right ].
\label{statuv}
\end{equation}
Here, $H_{sp}=-\frac{\hbar^2}{2m}\frac{\partial^2}{\partial x^2}
+\frac{1}{2}m\omega_x^2 x^2$ and $\mu$ is the chemical potential. 
The superscript $s$ denotes the static HFB
calculation, e.g. the static order parameter is denoted by $\Phi^{s}(x)$ 
and $n_c^{s}(x)=|\Phi^{s}(x)|^2$ denotes the static HFB condensate 
density \cite{quasicondensate}. The normal density is 
$\tilde n^{s}(x)=\sum_j {|u_j^{s}(x)|^2 N_j
+ |v_j^{s}(x)|^2 (N_j+1)}$ 
,  whereas $\tilde m^{s}(x)=-\sum_j
u_j^{s}(x)v_j^{s*}(x) (2N_j+1)$ is the anomalous density \cite{Griffin}. 
The population of excited states at temperature $T$ follows the Bose distribution 
$N_j=(e^{E_j/kT}-1)^{-1}$. In Eq. (\ref{statuv}) we denote 
$\mathcal{L}^{s}(x)=H_{sp}+2g_{1D}[n_c^{s}(x)+\tilde n^{s}(x)]-\mu$ 
and 
$\mathcal{M}^{s}(x)=-g_{1D}[\Phi^{s2}(x)+\tilde m^{s}(x)]$. 
Two variants of the HFB formalism are the Hartree-Fock (HF) approach, altogether
neglecting anomalous terms, and the HFB-Popov approximation wherein the 
{\it noncondensate} anomalous density is explicitly dropped, i.e.  
$\mathcal{M}^{s}(x)\approx -g_{1D}\Phi^{s2}(x)$ \cite{Griffin}.
Unlike the static HFB, the HF and HFB-Popov approximation 
do not have an unphysical gap in their excitation spectra \cite{Griffin}.

We use the solution of Eqs. (\ref{statcon}) and (\ref{statuv}) 
as the initial conditions to study dynamics without confinement 
within the TDHFB approximation in the modal form \cite{Rey}

\begin{equation}
i\hbar \frac{\partial \Phi(x,t)}{\partial t}=
H_{sp} \Phi+g_{1D}[n_c(x,t)+2\tilde n(x,t)]\Phi
+g_{1D} \tilde m(x,t)\Phi^*,
\label{con}
\end{equation}
\begin{equation}
i\hbar\frac{\partial}{\partial t}
\left [
\begin{array}{c}
u_j(x,t) \\
v_j(x,t) 
\end{array}
\right ]
=
\left [
\begin{array}{c c}
\mathcal{L}(x,t) &
\mathcal{M}(x,t) \\
-\mathcal{M}^{*}(x,t) & 
-\mathcal{L}(x,t)
\end{array}
\right ]
\left [
\begin{array}{c}
u_j \\
v_j 
\end{array}
\right ],
\label{dynuv}
\end{equation}
where 
$\mathcal{L}(x,t)=H_{sp}+2g_{1D}[n_c(x,t)+\tilde n(x,t)]$,  
$\mathcal{M}(x,t)=-g_{1D}[\Phi^{2}(x,t)+\tilde m(x,t)]$,
$\tilde n(x,t)=\sum_j {|u_j|^2 N_j + |v_j|^2 (N_j+1)}$ 
is the normal density,  and $\tilde m(x,t)=-\sum_j
u_jv_j^{*}(2N_j+1)$ is the anomalous density; at $t=0$, all dynamical 
quantities are identical to their static 
counterparts, e.g., $\Phi(x,0)=\Phi^{s}(x)$ etc. 
The TDHFB model is usually given in the form of coupled equations for 
the condensate order parameter and the single particle density matrix
(e.g. see Ref. \cite{Holland}). 
A tedious but straightforward calculation \cite{Derivation} shows its 
full equivalence to the modal form (\ref{dynuv}). As expected, 
when the time-dependence of the BEC and qusiparticle functions
is $\Phi(x,t)=\Phi^{s}(x)\exp(-i\mu t/\hbar)$, 
$u_j(x,t)=u_j^{(s)}(x)\exp(-i(E_j+\mu)t/\hbar)$, and 
$v_j(x,t)=v_j^{(s)}(x)\exp(-i(E_j-\mu)t/\hbar)$, the equations of motion
(\ref{con}) and (\ref{dynuv}) reduce to the time-independent
HFB equations (\ref{statcon}) and (\ref{statuv}).

In what follows we present numerical results based on the
described formalism, demonstrating the effect of thermal particles
and condensate depletion on matter-wave soliton dynamics. 
The parameters of the calculation are chosen to
resemble the experimental parameters of Ref. \cite{Khajkovic}. We
consider $N=2.2\ 10^4$ $^{7}$Li atoms in a harmonic trap with
$\omega_{\perp}=4907$ Hz ($a_{\perp}=\sqrt{\hbar/m\omega_{\perp}}\approx 1.35\ \mu$m).
The 3D scattering length $a_{3D}=-3.1\ 10^{-11}$ m corresponds to a
nonlinear parameter of $N|a_{3D}|\approx 0.68 \ \mu$m, and is tunable by the
Feshbach resonance technique \cite{Khajkovic}. 
We present two calculations for different $T$ and 
$\omega_x$ to show the influence of excitations on the 
soliton dynamics. 

\begin{figure}
\centering
\includegraphics[width=0.5\textwidth]{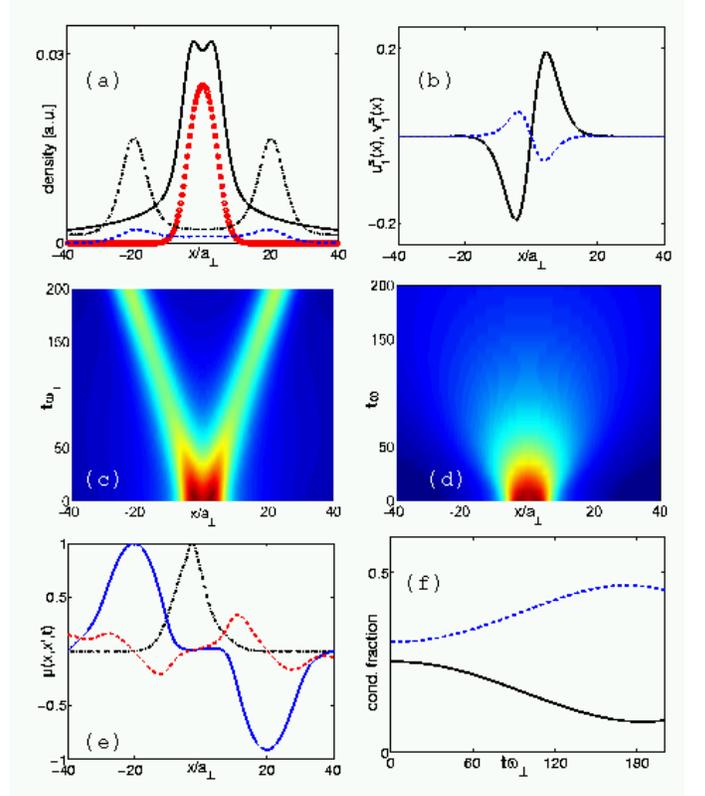}
\caption{(color online) (a) Total density (solid line), and 
condensate density (circles) at $t=0$; total (dot-dashed line)
and condensate density (dashed line) at $t\omega_\perp=180$. 
(b) The dipole-shaped first excited state, $u_1^s(x)$ (solid line)
and $v_1^s(x)$ (dashed line). 
(c) Evolution of the density with interactions present, 
and (d) without interactions. 
(e) The complex degree of coherence
$\mu(x,x',t)$ of a partially-coherent matter-wave
at $t=0$ (dot-dashed line), and $t\omega_\perp=180$ [$Re\ \mu(x,x',t)$ solid line, 
and $Im\ \mu(x,x',t)$ dashed line]; $x'$ is placed at the position of 
the left peak. 
(f) The condensate fraction (solid line) and the total population 
within the first 60 excited states (dashed line) during evolution. 
}
\label{fig1}
\end{figure}

First we consider the gas at higher temperature, where 
the condensate fraction is approximately $25$\%. 
The trapping frequency is 
$\omega_{\perp}=141$ Hz ($a_{x}=\sqrt{\hbar/m\omega_{x}}\approx 8.0\ \mu$m),
while $k_BT/\hbar \omega_{\perp}=60$.
Fig. \ref{fig1}(a) illustrates the total density
$n_c^{s}(x)+n_t^{s}(x)$ of the stationary HFB-Popov 
calculation. The total density profile is double humped, 
which is a clear signature of the significant population 
of the first excited state (approximately $\approx 10$\% of atoms), 
which has a dipole-like shape shown in Fig. \ref{fig1}(b). 
We have numerically checked the stability \cite{Kagan1} of the solution 
with respect to small perturbations; the stability is underpinned by 
the use of parameters resembling experiment \cite{Khajkovic}.
When the trap is turned off, the system is suddenly taken out of 
equilibrium, and consequently starts to evolve; we simulate the dynamics 
with the full TDHFB model. 
In the spirit of Ref. \cite{Khajkovic}, we compare the $x$-unconfined
dynamics of the system in the presence of interparticle interactions
[Fig. \ref{fig1}(c)] to its time evolution when both the confinement
in $x$ and the interactions are turned off [Fig. \ref{fig1}(d)].
In the absence of interactions, we clearly observe fast 
matter-wave dispersion. In contrast, when interactions are present, 
the two humps begin to separate, because the
trapping potential which provided a balance to the kinetic energy
term is no longer present [Fig. \ref{fig1}(c)]. 
During evolution, the condensate is slowly depleted 
[Fig. \ref{fig1}(f)]. Consequently, at one point during the evolution 
the two humps split, each forming a solitonic structure 
with opposite momentum. 
Each of these solitonic structures contains both Bose-condensed atoms, 
and a significant portion of the non-condensed particles, i.e., 
they are partially coherent matter-wave solitons. 
The uniqueness of such random-phase structures is ellucidated by 
their complex degree of coherence,
$\mu(x,x',t)=\rho(x,x',t)/\sqrt{\rho(x,x,t)\rho(x',x',t)}$,
where $\rho(x,x',t)=\Phi(x',t)^{*}\Phi(x,t)+
\langle\hat \Psi^{\dagger}(x',t)\hat \Psi(x,t)\rangle$, 
plotted at different times in Fig. \ref{fig1}(e).
We observe that spatial correlation is localized in space at 
$t=0$. Correlations change as the two humps split.
While retaining spatial localization, the phases at 
the two separated peaks are well-correlated, and out of phase, 
resembling the behavior of out-of-phase adjacent solitons 
from Ref. \cite{Strecker}.
We emphasize that for zero-temperature GPE solitons, the pair correlation function 
factorizes as $\rho(x,x')=\Phi^*(x)\Phi(x')$, which yields $\mu(x,x')=1$, 
corresponding to coherent matter-waves. 
Our incoherent matter-wave solitons are thus rather special in that
they correspond to localization of {\it entropy} and spatial correlation,
as well as to localization of density.

While the temperature $k_BT$ in the previous example is higher 
than the transverse level spacing $\hbar \omega_{\perp}$ whereas a 
'true' 1D geometry calls for $k_BT<\hbar \omega_{\perp}$ \cite{Moritz},
the use of a Q1D formalism is still justified because the first
$\omega_x/\omega_{\perp}\sim 35$ states are essentially 1D
(they are in the lowest state of the transverse Hamiltonian).
Furthermore, only condensed atoms, and atoms from lower excited 
states determine the outcome of the motion. Therefore, a proper inclusion of the
transverse dimension in the calculation would lead to some rescaling
of the parameters, but would not influence the dynamics 
observed in our quasi-1D calculation. Moreover, the simulations as well
as the experiment of \cite{Khajkovic}, are all in the weak interaction regime
$N|a_{1D}|/a_x\sim 10^8\gg 1$ \cite{Dunjko}, thus justifying the use 
of a mean-field approach.

\begin{figure}
\centering
\includegraphics[height=0.5\textwidth,angle=-90]{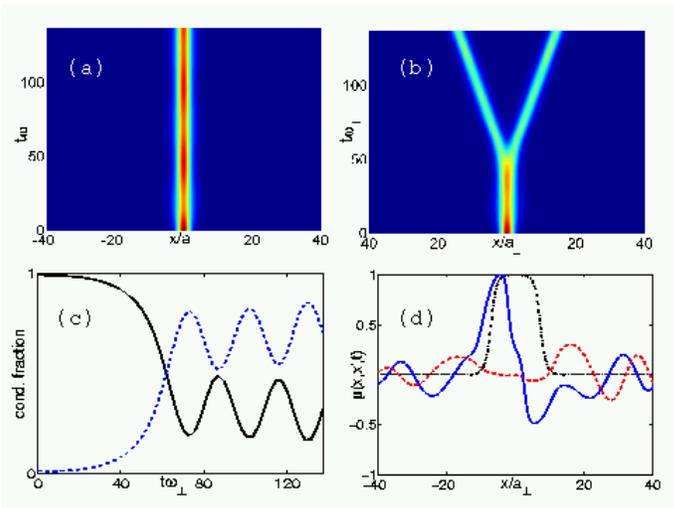}
\caption{(a) The evolution of the system with GPE, and 
(b) with TDHFB. (c) The condensate fraction (solid line) and 
the fraction of non-condensed atoms during TDHFB evolution. 
(d) The complex degree of coherence
$\mu(x,x',t)$ at $t=0$ (dot-dashed line), and 
after the condensate is considerably depleted 
at $t\omega_{\perp}=73$ [$Re\ \mu(x,x',t)$ solid line, 
and $Im\ \mu(x,x',t)$ dashed line]; $x'$ is placed at the position of 
the left peak.
}
\label{fig2}
\end{figure}

Next, we consider a gas prepared at near-zero temperature, where 
the condensate fraction is $99$\%. The trapping frequency is 
$\omega_{\perp}=439$ Hz ($a_{x}=\sqrt{\hbar/m\omega_{x}}\approx 4.51\ \mu$m),
while $k_BT/\hbar \omega_{\perp}=5$. As previously, the trap is turned
off and the dynamics is calculated using both the GP equation 
(Fig. \ref{fig2}a) and the TDHFB formalism (Fig. \ref{fig2}b).
The mechanical stability of the evolving BEC is demonstrated by 
propagating the time-dependent GP equation, showing small 
oscillations of the condensate width, rather than a mechanical 
collapse to a point [Fig. \ref{fig2}(a)]. This motion corresponds 
to a coherent matter wave soliton. However, as clearly evident from 
propagation of the TDHFB equations [Fig. \ref{fig2}(b)], allowing BEC 
depletion [Fig. \ref{fig2}(c)], the BEC collapses via a {\it pairing} 
instability \cite{Imry,Jeon} whereby pairs of atoms are collisionally 
pulled out of the BEC into the thermal cloud, thus gaining a mean-field
energy which goes into their relative motion. Such pairing collapse 
with little or no mechanical shrinking is indeed observed in 3D collapse 
experiments \cite{Donley,Milstein}. Our results show that in 1D 
it results in two separate solitonic structures of opposite momentum,
containing both a condensed part and a significant thermal population.
We note that similar structures were observed in stochastic simulations 
of molecular BEC dissociation in 1D geometry \cite{Kheruntsyan}, indicating
that incoherent matter-wave solitons can also be produced in this system.  
Figure \ref{fig2}(d) shows that the correlations of each solitonic 
structure are localized, and the two structures are partially 
correlated.

Before closing, we note that within the time-dependent 
Hartree-Fock (TDHF) approximation \cite{Proukakis1}, where condensate depletion 
is not accounted for during the dynamics, we find solutions 
with double-humped density profile that begin to split 
as in TDHFB model. However, as the condensate is not depleted within TDHF,
the two peaks may be pulled back and merge to almost recover the initial 
density profile. Such motion is also characteristic of composite incoherent 
solitons in optics.

In conclusion, we have used the time-dependent Hartree-Fock-Bogoliubov
theory to analyze the influence of the thermal cloud and condensate 
depletion onto the dynamics of BEC solitons. We find that 
condensate depletion induced by pairing, and the presence of a thermal 
cloud cause the particle density to split into two solitonic structures,
each being a mixture of condensed and non-condensed particles.
The predicted incoherent matter-wave structures represent novel correlation
solitons which resemble localized second-sound entropy waves.
Such random-phase matter-wave solitons correspond  
to incoherent solitons in nonlinear optics \cite{Mitchell,Equ,Mitchell1}, 
which points at the analogy between partially condensed Bose gases and nonlinear 
partially coherent optical waves.

\end{document}